\documentclass{article}
\usepackage{amsmath}
\usepackage{booktabs}
\usepackage{multirow}
\usepackage{graphicx}
\usepackage{siunitx} 
\usepackage{float}
\usepackage{subfig}
\usepackage{url}

\title{Superstatistical Analysis of PDFs and autocorrelation functions for air pollution concentrations in the UK}

\author{Nisal Amarakoon$^1$, Hankun He$^2$$^3$, Christian Beck$^1$}

\date{$\,^1$ Centre for Complex Systems, Queen Mary University of London, London E1 4NS, UK \\ $^2$ UK Centre for AI in the Public Sector, London, United Kingdom \\ $^3$ Institute for Connected Communities, University of East London, London E15 4LZ, United Kingdom}

\begin{document}

\maketitle
\abstract{{\bf Conventional statistical models often struggle to fully capture the complex spatio-temporal dynamics, intermittent fluctuations, and heavy-tailed distributions characteristic of real-world air pollution data. Furthermore, existing literature frequently focuses on extreme events, overlooking the persistence of low-pollution states and temporal memory effects. To address these gaps, we apply superstatistical frameworks from non-equilibrium statistical physics to analyse a comprehensive five-year dataset (2020–2025) of hourly air pollutant concentrations across the United Kingdom. Excellent fits of experimentally measured distributions are obtained from our theoretical models. We observe large heterogeneities of the best fitting parameters depending on the locations where the measurements are performed. These parameters 
 form characteristic patterns in the 3-dimensional parameter space and depend on the type of pollutant considered, as well as on the environmental conditions (high traffic, industrial, or rural surroundings). We also investigate autocorrelation functions and provide evidence for differences in day-time and night-time decays of the autocorrelation function. Our investigation mainly focuses onto the dynamics of $NO, NO_2$, $PM_{2.5}$, $PM_{10}$, but we also report on some anomalous distributions observed for $O_3$.}}

\section*{Impact Statement}
Our application paper offers a substantial research contribution at the interface of statistical physics, environmental sciences, geography, and data analytics. It addresses the complex dynamics of air pollution in UK,
analyzing, in particular, the tails of the observed PDFs which describe high-pollution events as well as the low pollution data. Our work enables a
better understanding of the time-varying dynamics of air pollution, which is essential for policy formulation and
the construction of suitable stochastic models, as well as for analyzing the medical consequences of exposure to
polluted air.

\section{Introduction}

Air pollution represents one of the most pressing environmental challenges of our time, with far-reaching consequences for human health, ecosystems, and climate systems \cite{Seigneur2019}. Hill emphasizes that “Environmental scientists rely heavily on statistical methods to interpret data, discern patterns, and assess the risks associated with pollutant exposure” \cite{Hill2010}. Conventional statistical approaches, while useful for basic analysis, often prove inadequate in characterizing the intricate spatio-temporal patterns and complex distributions observed in real-world pollution data. The field has witnessed significant theoretical advancements through the application of generalized statistical physics principles, particularly through frameworks such as superstatistics \cite{Beck2003,Beck2007,Metzler2020,Williams2020,He2022}
and nonextensive statistical mechanics \cite{Tsallis1988, Tsallis2004,Tsallis2023}. These methodologies provide robust tools for analysing complex pollution dynamics, including heavy-tailed non-Gaussian probability distributions, long-range correlations, {{intermittent fluctuations, and region-specific variations that defy traditional modelling assumptions. Our paper here discusses the evolution of air pollution statistical descriptions from $q$-exponential distributions \cite{Tsallis1988} to $q$-Gamma formulations, which take into account both high pollution and low-pollution asymptotics, highlighting their respective strengths in addressing different aspects of environmental characteristics. Through careful examination of both theoretical foundations and practical applications, our work underscores how physics-inspired statistical techniques offer superior capabilities for understanding the typical behaviour of pollution probability density functions (PDFs),
allowing for pollution statistical predictions, environmental risk assessment, and evidence-based policy formulation. The integration of these advanced methods into air quality research marks a significant step forward in our ability to understand and mitigate the impacts of the fluctuating aspects of atmospheric pollution.


The quantitative analysis of air pollution statistics began with early applications of parametric probability distributions, illustrated by Marani et al. \cite{Marani1986} demonstrating that generalized gamma distributions could effectively model air pollutant concentrations. This foundational study established the importance of statistical methods in air quality research and addressed the need for flexible distributions that could accommodate the right-skewed nature of pollution data. While innovative at its time, this approach was ultimately limited by its assumption of stationary conditions and global statistical uniformity - constraints that would later motivate more sophisticated modelling frameworks.

More recently, Williams et al.\cite{Williams2020} develop a novel framework for analysing air pollution dynamics through the lens of superstatistics, by now a standard method in nonequilibrium statistical mechanics \cite{Beck2003, Beck2007, Metzler2020}. This approach, rooted in statistical physics, addresses the limitations of conventional models by treating pollution fluctuations as arising from a superposition of different statistical regimes. The authors demonstrate how this method can capture the heavy-tailed probability distributions and intermittent behaviour characteristic of real-world pollutant concentration data. By incorporating time-scale separation between rapid fluctuations and slower environmental variations, their model provides improved representation of extreme pollution events that often evade traditional Gaussian or log-normal descriptions. This work establishes a crucial theoretical foundation for understanding the complex temporal patterns in air pollution, particularly in urban settings where multiple emission sources interact with changing meteorological conditions.

Expanding beyond temporal analysis, He et al.\cite{He2022} examine the geographical dimension of pollution variability, in their case for a large European data set. The research reveals significant regional differences in pollution statistics across Europe, challenging the assumption of spatial uniformity in air quality modelling. Through advanced spatial analysis techniques, the authors identify distinct statistical signatures in pollution data that correlate with factors such as local emission sources, topographic features and climate patterns. This work provides empirical evidence that pollution dynamics cannot be adequately described by universal constant parameter statistical models, but rather require region-specific approaches. Together, these two studies offer a comprehensive perspective on air pollution statistical features, integrating both temporal and spatial complexity through innovative applications of statistical physics principles to environmental science.

Autocorrelation analysis is another core tool in understanding the persistence and predictability of air pollutant concentrations. The authors of \cite{golubnichiy2020} investigated PM$_{10}$ concentrations in the Krasnoyarsk Territory using autocorrelation analysis to explore temporal patterns in particulate matter data from atmospheric monitoring systems. Their work highlights the presence of short term and long term dependencies in PM$_{10}$ time series, which can inform forecasting models and pollution control measures. By focusing on a single pollutant within a specific regional monitoring framework, the study provides valuable insights into local air quality dynamics, while also illustrating the broader need to address both the temporal structure and environmental drivers behind pollutant fluctuations.

Expanding beyond a single location and pollutant, Dai and Zhou \cite{dai2017} analysed temporal and spatial correlation patterns across Chinese cities using hourly monitoring data and network-based approaches such as the Planar Maximally Filtered Graph method. Their findings reveal pollutant specific behaviours, such as strong spatial correlations in ozone and more localised dispersion patterns in particulate matter, and show the persistence of pollution events through long term temporal memory. The identification of geographically coherent clusters of cities with synchronised pollutant dynamics provides an important framework for regional pollution control, emphasizing that policy interventions must account for both meteorological influences and industrial activity patterns across space and time.

A key gap in the literature dealing with statistical fluctuations of air pollutant concentrations is the lack of focus on low pollution periods (besides the high-pollution states that are often investigated \cite{eds}) and also the discussion of memory effects, i.e. how pollution levels depend on their past values over time, and how correlation functions decay. Most studies concentrate on high pollution events and extreme cases. However, even during low-pollution periods, pollution can show clear persistence due to weather conditions and ongoing background emissions \cite{Yu2022AirPollution}. Ignoring this can lead to incomplete descriptions of the dynamical behaviour. This illustrates the need for models that also capture the PDF behaviour for low pollution situations, which will lead to the more general PDFs studied in this paper, exhibiting power-law behaviour both for high- and low-pollution situations. In addition to this, we will also study correlation functions of measured pollution concentrations, describing the dynamical behaviour.

\section{Methodology}

In this paper, we analyse a large amount of data measured between 2020 to 2025 in the UK, and fit the observed non-Gaussian probability density functions (PDFs) of measured air pollution concentrations  with functional forms that are motivated by superstatistical models \cite{Beck2003,Beck2007} and the formalism of non-extensive statistical mechanics \cite{Tsallis1988,Tsallis2023}, i.e.\ in general by methods borrowed from the physics community when describing complex systems in nonequilibrium states. We will  provide a systematic investigation how the best-fitting parameters for the observed PDFs vary at different spatial locations and for different types of pollutants, forming characteristic clouds and patterns in the parameter space. 

\begin{figure}[H]
    \centering
    \includegraphics[width=.98\linewidth]{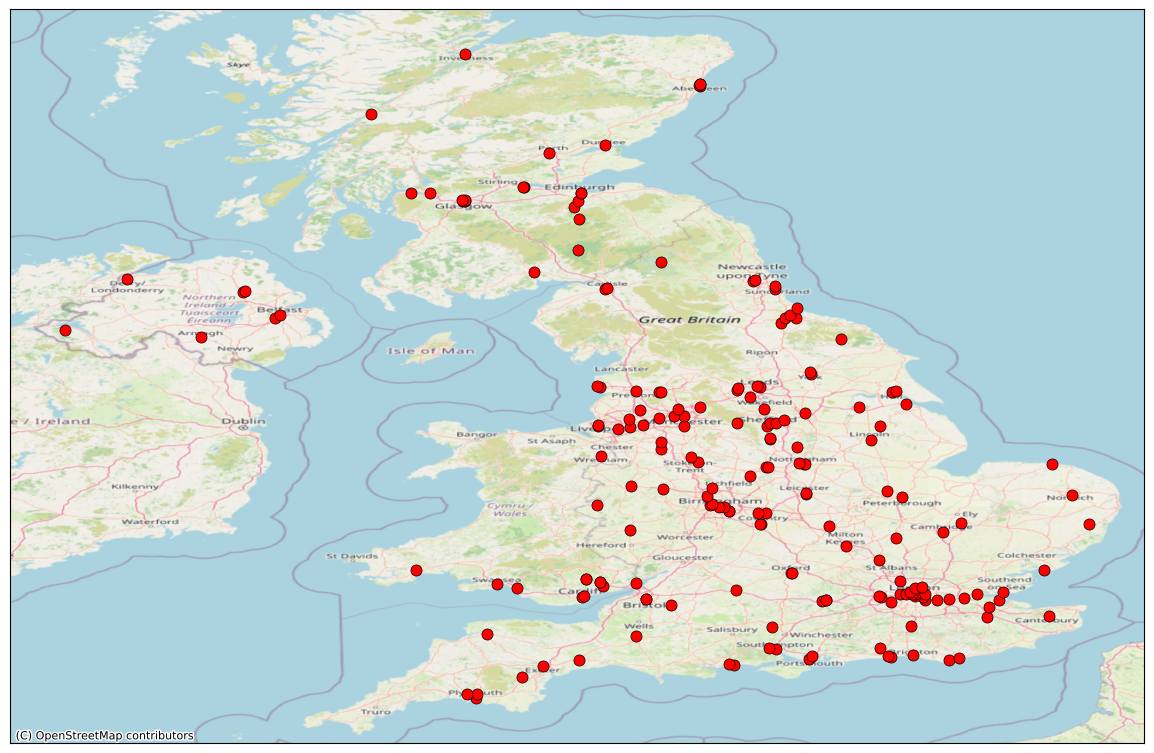}
    \caption{Geographical distribution of UK air quality monitoring sites analysed in this paper. The map illustrates the locations of monitoring stations (red markers) across the United Kingdom, as recorded by the Department for Environment, Food and Rural Affairs (DEFRA).}
    \label{map}
\end{figure}

The spatial data utilized in this study was sourced from the UK-AIR database, comprising hourly concentration measurements recorded between January 2020 and January 2025 across all available monitoring sites in the United Kingdom at this length of time \cite{UKAirDefra}. To evaluate spatial heterogeneity, these sites were categorized into distinct environmental classifications, including Urban Traffic, Urban Background, Urban Industrial, Suburban Industrial, Suburban Background, and Rural Background. The statistical analysis primarily focused on characterizing the distributions of four key atmospheric pollutants: particulate matter 2.5 ($PM_{2.5}$), particulate matter 10 ($PM_{10}$), nitric oxide (NO), and nitrogen dioxide ($NO_2$). While Ozone ($O_3$) was also initially investigated, it frequently exhibited anomalous probability density functions and was therefore excluded from the primary modelling. Prior to the full parameter space analysis, the empirical data was filtered to take positive values only (as expected for a concentration); also a few specific sites that yielded atypical distributions and failed to produce a good fit with the theoretical models were discarded, most of these discarded sites lacked a significant amount of data. The remaining observed probability density functions were fitted with $q$-Gamma distribution. To determine the specific distribution parameters ($\alpha$, $q$, and $\lambda$), the  Maximum Likelihood Estimation (MLE) parameter estimation technique was employed, which identifies values that maximize the probability of the observed data being produced by the given distribution with those parameters.


The probability density functions (PDFs) that we will use in our novel fitting approach in the following are the so-called \textbf{$q$-Gamma distribution}s, defined as
\begin{equation}
f(x) = \frac{1}{Z} \, x^{\alpha-1} \left[ 1 + (q-1)\lambda x \right]^{\frac{1}{1 - q}},
\end{equation}
where
\begin{itemize}
    \item $\alpha$ controls the shape of the probability density at low concentrations $x$ ($x \to 0$),
    \item $q$ quantifies the non-extensiveness (deviation from classical exponential decay obtained for the special case  $q\to 1$),
    \item $\lambda$ is an inverse scale parameter,
    \item $Z$ is a normalization constant. 
\end{itemize}
These types of distributions occur in generalized statistical mechanics models of complex systems (see, e.g.\ \cite{Tsallis2023} for a recent comprehensive review), as well as in superstatistical modelling approaches, where the parameters of a simple dynamical model (such as local stochastic differential equation) are random variables itself, for details, see e.g.\ \cite{Beck2007}. The relevance of superstatistical models for air pollution dynamics has been previously mentioned and applied to the high pollution tail behaviour of observed PDFs, see \cite{He2022, eds}. Here we go a step further, by also incorporating low-pollution events, thus taking into account the entire range of possible pollution states, from very low to very high.

\section{Results}

\subsection{Observed PDFs--typical examples} 
We focus our statistical analysis primarily onto the following  air pollutants: particulate matter
2.5 ($PM_{2.5}$), particulate matter 10 ($PM_{10}$), nitric oxide ($NO$) and nitrogen dioxide ($NO_{2}$), investigating their dynamics for all available sites in the United Kingdom as recorded on \cite{UKAirDefra}. We have investigated the Probability density functions (PDFs) of air pollution concentrations at all these locations and checked whether our theoretical model PDFs yield a good fit. The vast majority of sites are fitted well, though with different values of the parameters.

We start with an illustrative example of a typical trajectory of an air pollution concentration over time, as shown in Fig.~\ref{NO2_con_v_time}.
\begin{figure}[H]
    \centering
\includegraphics[width=1\linewidth]{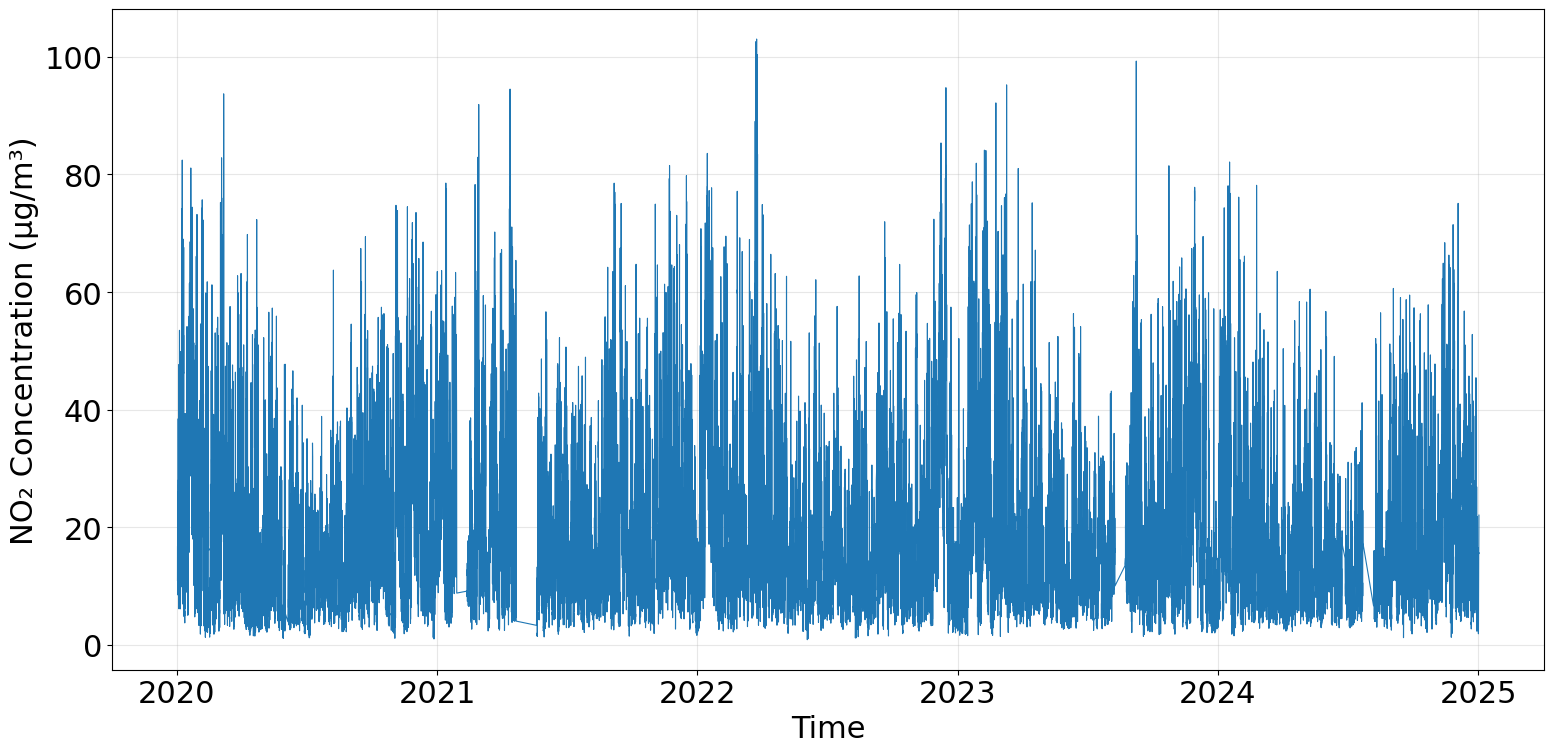}
\caption{The time series trajectory plot of $NO_{2}$ concentration for the Leicester location between 2020 and 2025, measured at 1 hour intervals.}
\label{NO2_con_v_time}
\end{figure}
The graph of Fig.~\ref{NO2_con_v_time} shows how $NO_{2}$ levels meassured at an example site in Leicester vary over time. Pollution levels fluctuate across multiple temporal scales, with occasional sharp concentration spikes observed. These spikes are likely caused by special short pollution events, such as high traffic or certain weather conditions. There is also a clear long-term oscillating pattern that repeats over the years, suggesting that $NO_{2}$ levels are also affected by seasonal patterns.

In the following, we look at the PDFs generated for various types of pollutants. The use of linear-linear, log-linear, and log-log scales in Fig.~\ref{Allpdfs} 
allows us to see which regions of the PDFs are particularly well fitted (for example those around the tails or those around the maximum). Linear-linear plots characterise the "bulk" of typical data, but they often do not yield much information on the behaviour of rare events in the tails. Log-linear (y-log) scales visualise exponential decay as a straight line, allowing for a precise assessment of how quickly concentrations diminish from the mean. Crucially, log-log plots are useful to identify power-law behaviour as straight lines. 

\begin{figure}[H]
    \centering
\includegraphics[width=1\linewidth]{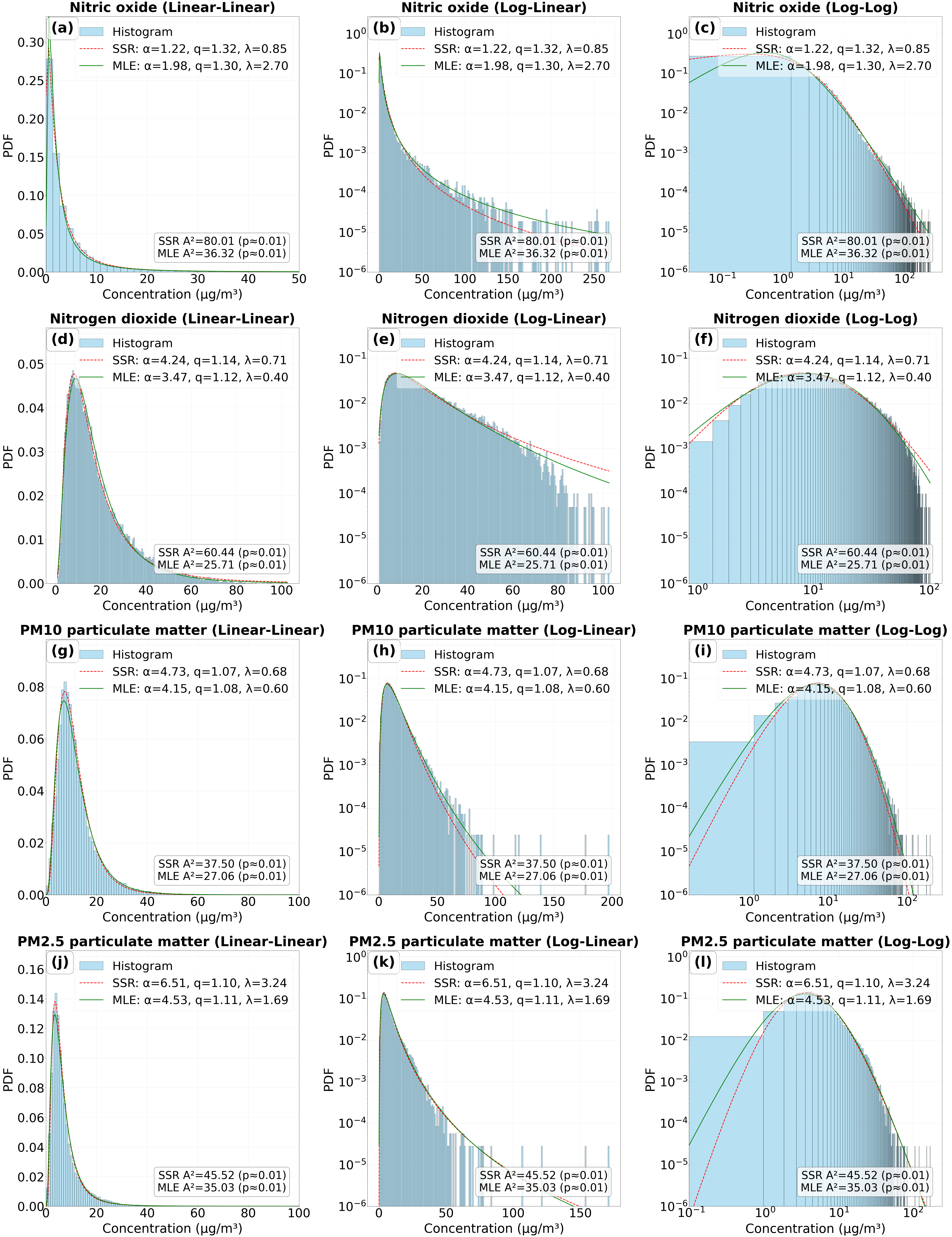}
\caption{PDFs of air pollutant concentrations for the example site at Leicester. Each row displays the distributions of a different pollutant:$NO$, $NO_{2}$, $PM_{10}$ and $PM_{2.5}$. To highlight different characteristics of the tail behaviours, the columns present the same data across three different axis scales: Linear-Linear (left), Log-Linear (middle), and Log-Log (right). The empirical data is represented by light blue histograms. Theoretical curves are $q$-Gamma fits using two different estimation methods: Sum of Squared Residuals (SSR, red dashed lines) and Maximum Likelihood Estimation (MLE, green solid lines).}
\label{Allpdfs}
\end{figure}

The PDF shown in the 2nd row of Fig.~\ref{Allpdfs} actually corresponds precisely to the trajectory shown in Fig.~1,

The distribution of Nitric Oxide ($NO$) (Fig.~\ref{Allpdfs} a-c) represents the most extreme case of heavy-tailed behaviour among the 4 investigated pollutants. The non-extensiveness parameter is consistently high ($q=1.30$ for MLE and $q=1.32$ for SSR), indicating a power-law-like decay that persists even at very high concentrations. This suggests that the $NO$ time series is characterised by frequent high-intensity "spikes," likely due to its proximity to primary combustion sources. In contrast, Nitrogen Dioxide ($NO_2$) (Fig.~\ref{Allpdfs} d-f) shows a more moderated distribution with $q$ values closer to unity ($q=1.12$ for MLE; $q=1.14$ for SSR). While $NO_2$ still exhibits heavy-tailed characteristics, the decay is markedly faster than that of $NO$. Quantitatively, the MLE method proves superior for both cases, achieving an Anderson-Darling ($A^2$) statistic of 36.32 for $NO$ and 25.71 for $NO_2$, significantly outperforming the SSR fits in both instances.

The final two rows of Fig.~\ref{Allpdfs} (subplots g-l) depict the distributions for $PM_{10}$ and $PM_{2.5}$. Both pollutants demonstrate a transition toward near-exponential behaviour, as evidenced by $q$ parameters approaching 1. For $PM_{10}$, the MLE and SSR fits are nearly identical ($q=1.08$ and $1.07$, respectively), while $PM_{2.5}$ shows slightly higher values ($q=1.11$ for MLE and $1.10$ for SSR). This difference suggests that while particulate matter concentrations are susceptible to occasional heavy-tailed events, they lack the extreme stochastic volatility observed in $NO$. Notably, the scaling parameter $\lambda$ varies significantly between the two, with $PM_{2.5}$ showing a much higher MLE value ($\lambda=1.69$) than $PM_{10}$ ($\lambda=0.60$). Consistent with the two pollutants, the MLE approach remains the more statistically sound methodology for these particulates, yielding lower $A^2$ values ($27.06$ for $PM_{10}$ and $35.03$ for $PM_{2.5}$) compared to the SSR approach.

\begin{figure}[H]
    \centering
\includegraphics[width=1\linewidth]{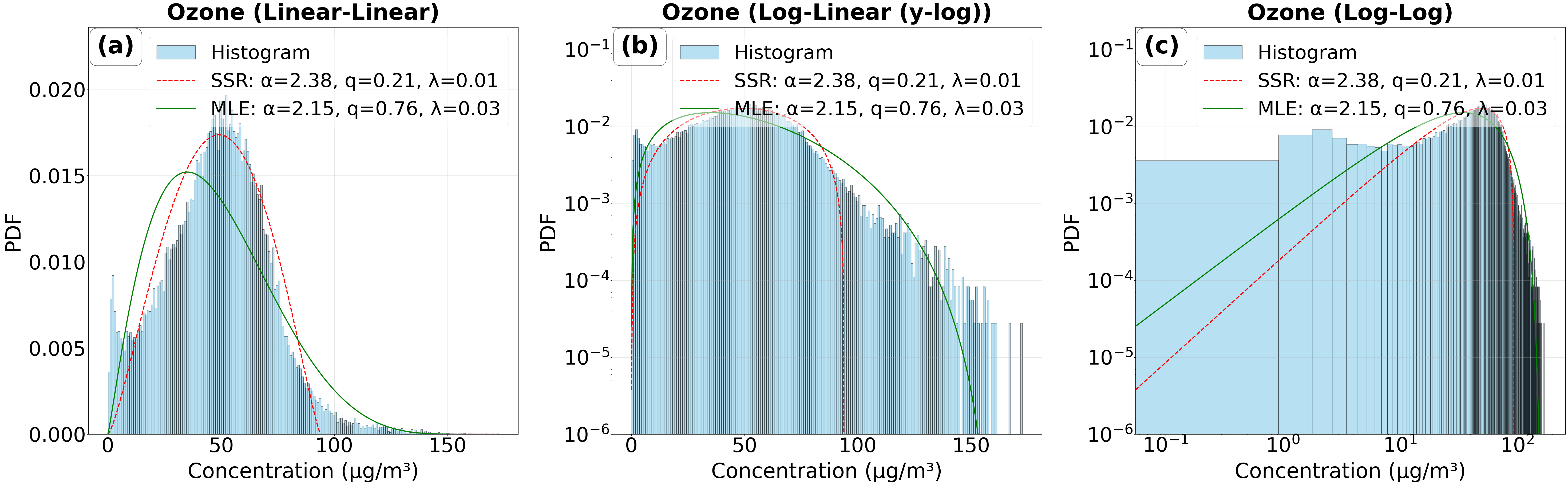}
\caption{Same as Fig.~\ref{Allpdfs}, but for Ozone ($O_{3}$).}
\label{ozone_pdfs}
\end{figure}

While the $q$-Gamma distribution captures 
quite efficiently the behaviour of the $NO_x$ and $PM_x$ pollutants discussed so far, there are substances with anomalous behaviour. An example is Ozone ($O_{3}$), typically observed to have PDFs that are quite different from those discussed so far. The reason may be the existence of decay modes for $O_3$.  Fig.~\ref{ozone_pdfs} illustrates that the $q$-Gamma struggles to accurately model Ozone concentrations at the Leicester site (as well as at other sites as well). The $q$-Gamma does not capture the 2 maxima observed. The Anderson-Darling test also fails to give us a goodness of fit test statistic.
This highlights the need for further methodological development to account for the distinct statistical behaviour of Ozone concentrations in future studies.

\subsection{Parameter space of $q$-Gamma fits}

The parameter relationship plots of Fig.~\ref{Parameter_relation} for all pollutants 
measured at the sites displayed in Fig.~1 reveal consistent clustering patterns across different site types, reflecting distinct emission and mixing characteristics. Only MLE is used for the fittings, as it has been shown to be the better fitting method in the previous section.

Urban Traffic sites consistently exhibit the highest $q$-values (often between $1.1$ and $1.3$) across all pollutants, paired with moderate $\alpha$ values (typically $2 – 6$). This indicates strongly non-extensive behaviour and heavy-tailed concentration distributions, characteristic of direct, intermittent emission sources such as vehicle exhaust.

Urban Background and Urban Industrial sites form an intermediate cluster, with $q$-values ranging from around $1.0$ to $1.3$ and $\alpha$ values spanning a wider range (approximately $2 – 8$). This suggests more varied and mixed emission sources, including traffic, heating, and industrial contributions, leading to less extreme but still heavy-tailed concentration profiles.

Suburban and Rural Background sites display the lowest $q$-values (generally $1.0 – 1.15$) and the highest $\alpha$ values (often $4 – 10$ or more), particularly evident in the longer-tailed plots (right-hand panels). This reflects smoother, more well-mixed concentration distributions typical of regional background air, with less influence from nearby strong emission sources.

The observed parameter differences between site  environments are statistically significant and indicate that air pollutant concentrations, in their dynamical behaviour, are strongly influenced by the location where they are measured. Traffic sites, understandably, have extreme values. Rural background sites have much more moderate values. Urban and industrial sites have values in between.

\begin{figure}[H]
    \centering
    \includegraphics[width=.98\linewidth]{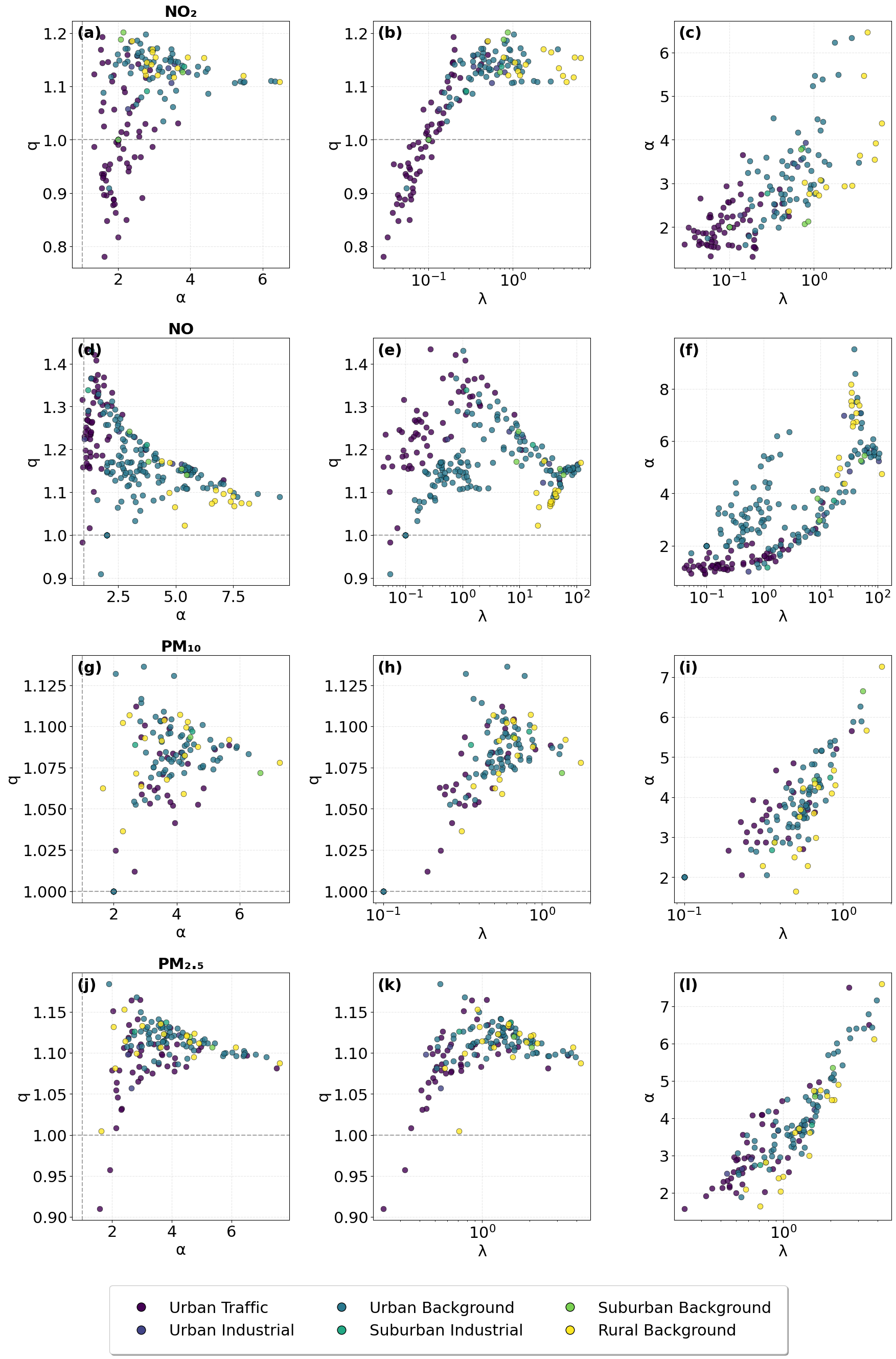}
    \caption{Parameter relationship plot of $q$ vs $\alpha$, $q$ vs $\lambda$ and $\alpha$ vs $\lambda$ fitted by $q$-Gamma for $NO_{2}$, $NO$, $PM_{10}$ and $PM_{2.5}$ time series. Each dot corresponds to a particular location in the data base.}
    \label{Parameter_relation}
\end{figure}

\subsection{Temporal correlation functions} 

By calculating the autocorrelation, we can distinguish between pollutant concentrations that dissipate rapidly and those that linger for longer. We determined temporal correlation functions
$C(t)=(E(x(s)x(s+t) -E(x(s)^2))/E(x(s)^2)$ from the measured air pollution time series $x(s)$. The correlation functions tend to decay slower than exponential. We fitted
these decays with $q$-exponential functions 
$e_q^{-\lambda t}= (1+(q-1)\lambda t)^{\frac{-1}{q-1}}$ (reducing to exponentials for $q=1$). 
Correlation functions of this type can by produced by superstatistical paramter fluctuations of the parameter $\lambda$ in an exponential decay $e^{-\lambda t}$. Some results,
for the example of the site London Marylebone Ropad, are shown in the following. 
\begin{figure}[H]
    \centering
    \includegraphics[width=1\linewidth]{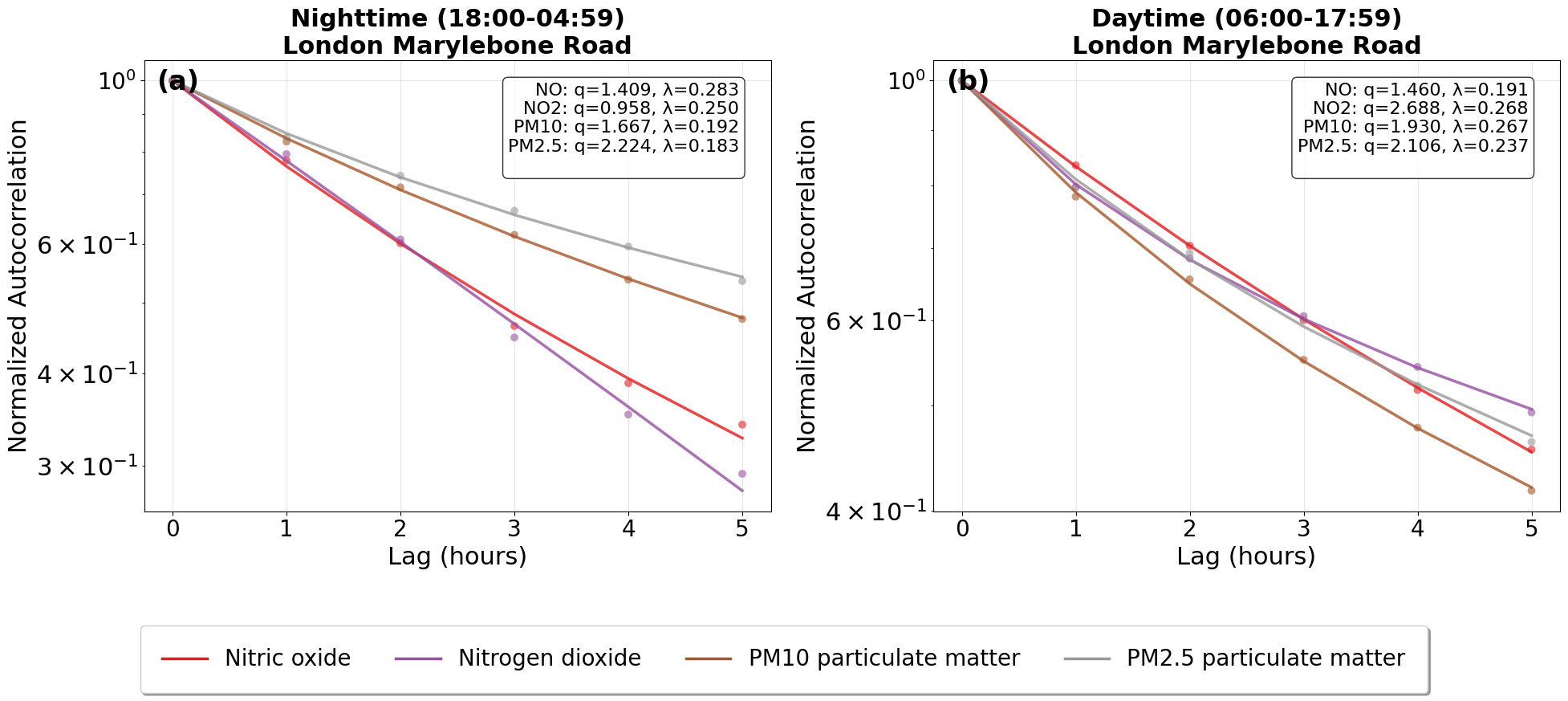}
    \caption{(a) - Daytime autocorrelation function at 5 lags fitted with $q$-exponentials for $NO$, $NO_{2}$, $PM_{2.5}$ and $PM_{10}$ time series measured at London Marylebone Road. (b) - Same as in (a) but for nighttime.}
    \label{Auto_q_expo}
\end{figure}

The $q$-exponential fits reveal distinct temporal correlation patterns across all pollutants and exhibit different behaviour if conditioned on either day- or night-time. During daytime, all pollutants exhibit strong non-extensive behaviour with $q$-values significantly greater than $1$ ($q=1.460-2.688$), indicating long-range correlations and heavy-tailed decay in autocorrelation. $NO_{2}$ shows the most extreme non-extensiveness ($q=2.688$), suggesting particularly persistent concentration patterns, while particulate matter demonstrates intermediate effects. At night, the correlation structure undergoes dramatic changes, nitrogen dioxide transitions to near-exponential decay ($q=0.958$), while nitric oxide maintains non-extensiveness ($q=1.409$) and particulate matter increases in correlation strength ($PM_{2.5}$: $q=2.224$). The $\lambda$ parameters, representing the initial decay rate for small time lags, show consistent values across pollutants ($0.183-0.283$) with nitric oxide displaying the fastest correlation decay at night ($\lambda=0.283$) and nitrogen dioxide the slowest during day ($\lambda=0.268$).

\begin{figure}[H]
    \centering
    \includegraphics[width=1\linewidth]{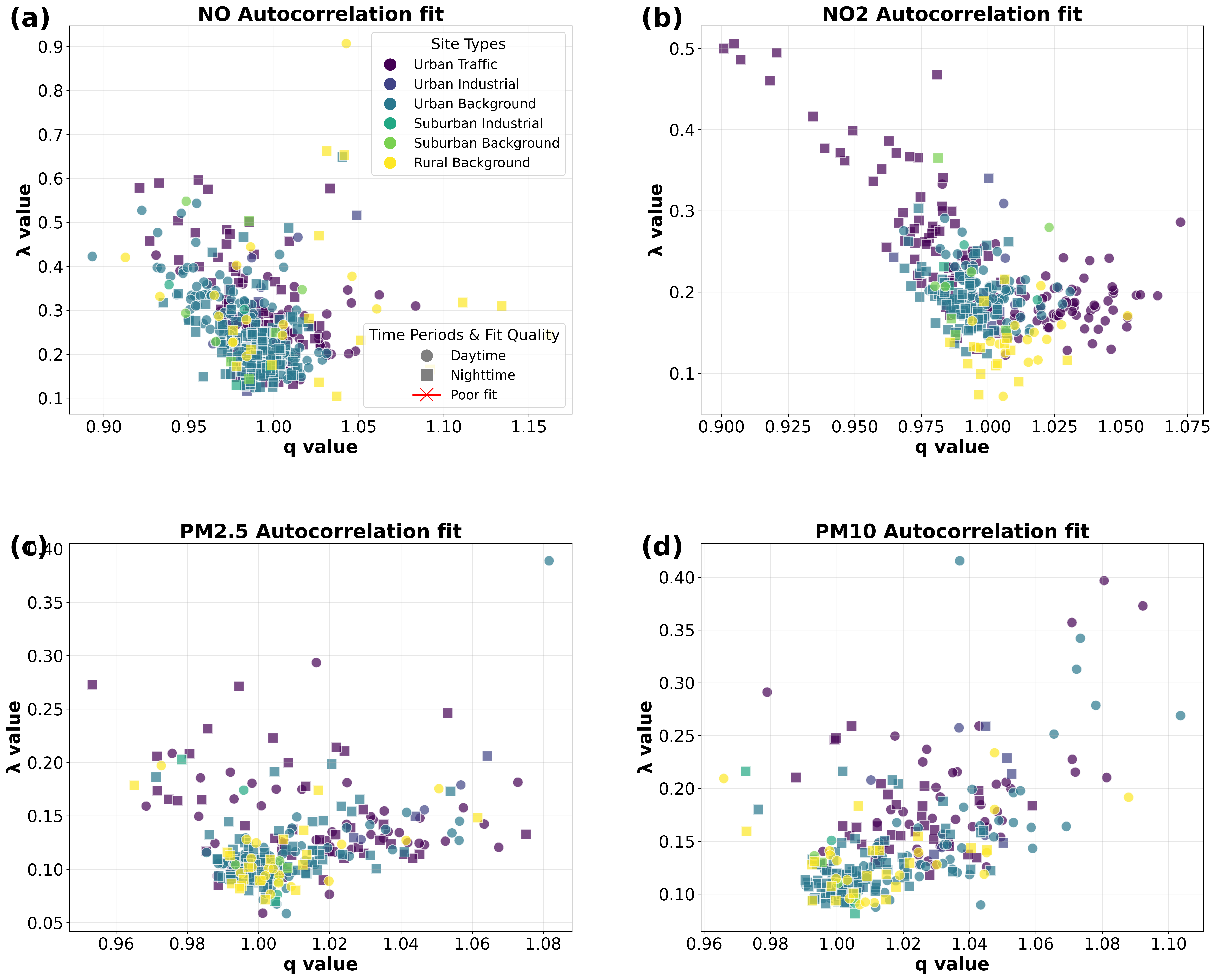}
    \caption{Best fitting parameters $q$ and $\lambda$ for measured shape of auto correlation function of $NO$, $NO_{2}$, $PM_{2.5}$ and $PM_{10}$ time series.}
    \label{q_l_scatterplot}
\end{figure}

Again we may proceed to the parameter space,
which for the correlation function fits is just
the 2-dimensional $(q,\lambda)$ plane. Our results for the best fitting parameters $(q,\lambda)$ at the various locations displayed in Fig.~1 are shown in Fig.~\ref{q_l_scatterplot}. 
As shown in the four scattering plots for (a) $NO$, (b) $NO_{2}$, (c) $PM_{2.5}$, and (d) $PM_{10}$, the fitted $q$ and $\lambda$ values for all pollutants are concentrated in broad, overlapping clusters. There is no clear visible influence of different site types or between daytime and nighttime periods across the plots.

In all plots the $q$-values cluster around 1.0, with most data points concentrated between approximately 0.95 and 1.10. This central clustering suggests that, across different locations and times, the underlying concentration dynamics for these pollutants often approximate a system near classical, extensive statistics. The $\lambda$ values associated with these 
$q$-values vary widely, ranging from about 0.1 to 0.8, indicating a broad range in the strength of temporal correlation.

There is no strong functional relationship between $q$ and $\lambda$ visible for (a),(c), and (d). However, plot (b) for nitrogen dioxide ($NO_{2}$) provides evidence for a pattern where the fitted $\lambda$ value appears to decrease as the $q$-value increases. This downward trend is most visible for points fitted during the daytime, suggesting that for daytime $NO_{2}$, stronger non-extensive behaviour correlates with weaker decay of the temporal correlation. In contrast, plots (a), (c), and (d) for NO, $PM_{10}$, and $PM_{2.5}$ show widely intermixed points for daytime and nighttime, with no pronounced systematic trend in the parameters. Although urban site types may show a slight tendency towards lower $\lambda$ values in some of the plots, there is substantial overlap of the corresponding background sites. 
Note that for $NO$ and $NO_2$ the average $q$-value is roughly equal to 1, whereas for $PM2.5$ and $PM10$ it is bigger than 1.

\section{Conclusion}

Based on the analysis presented in this paper, the $q$-Gamma distribution has proven to be a powerful tool for modelling PDFs of temporally varying air pollution concentrations, effectively capturing both, low-level and extreme high-concentration events. Unlike traditional models, this accounts for the observed heavy-tailed nature of pollutant concentrations, leading to more accurate descriptions of air pollution PDFs. Our fittings were quite successful for $NO, NO_2,PM_{2.5}$ and $PM_{10}$. However, 
there are limits of the applicability of this approach. For example, for Ozone we observe other types of distributions. Thus, the complexity of air pollution PDFs is inherently depending on the type of substance considered, suggesting that future research must explore alternative or multi-modal superstatistical frameworks to capture the dynamics of these anomalous pollutants.

Our analysis also reveals, by analysis of the parameters underlying the the PDF and autocorrelation structure,  that different types of monitoring sites, such as urban traffic, industrial and rural backgrounds, exhibit different statistical signatures that reflect their unique emission sources and environmental conditions. Additionally, the autocorrelation analysis shows clear differences between the dynamics of pollution at night and daytime, with pollutants such as $NO_{2}$ showing strong persistence during the day but near-exponential decay at night. These distinct spatio-temporal signatures show the need of region-specific and time-sensitive approaches to air quality management, demonstrating that policy interventions must account for local environmental drivers rather than relying on global, universally constant assumptions.

To summarize, our systematic spatio-temporal analysis of air pollution data for the UK shows that $q$-Gamma distributions and $q$-exponential correlation functions provide an excellent modelling framework for generic air pollution dynamics. These types of distributions arise naturally from superstatistical models, i.e. from stochastic differential equations where the parameters themselves are random variables.
These types of models provide a more subtle understanding of variations in pollution behaviour, supporting better environmental monitoring, risk assessment, and evidence-based policymaking, aimed at mitigating the impacts of extreme atmospheric pollution. As shown in this paper,
our parametrization in terms of the parameters $q$, $\lambda$ and $\alpha$ can help to quantify and comnpare the statistical properties of different pollutants at a variety of different locations.

\section{Appendix}

\subsection{Normalization Factor}

The normalization constant $Z$ is
\[
Z = \int_0^\infty x^{\alpha-1} \left[ 1 + (q-1)\lambda x \right]^{\frac{1}{1-q}} \, dx
\]
Using the substitution $v = (q-1)\lambda x$,
\[
Z = \frac{1}{[(q-1)\lambda]^\alpha} \int_0^\infty v^{\alpha-1} (1 + v)^{\frac{1}{1-q}} \, dv
\]
this evaluates to
\[
Z = \frac{\Gamma(\alpha) \, \Gamma\left( \frac{1}{q-1} - \alpha \right)}{[(q-1)\lambda]^{\alpha} \, \Gamma\left( \frac{1}{q-1} \right)}
\]

The $q$-Gamma distribution captures atmospheric pollution complexities through:
\begin{itemize}
    \item \textbf{the power-law term} ($x^{\alpha}$) $\rightarrow$ low-level pollution events,
    \item \textbf{the $q$-exponential term} (heavy tails) $\rightarrow$ extreme high-concentration events.
\end{itemize}

\subsection{Parameter Insights for Environmental Science}

\begin{align*}
    q < 1 &\rightarrow \text{Rapid decay (faster than exponential): confined or short-range processes}, \\
    q \approx 1 &\rightarrow \text{Classical exponential decay}, \\
    q > 1 &\rightarrow \text{Heavy-tailed decay (slower than exponential)}, \\
    \alpha > 0 &\rightarrow \text{Controls asymmetry; useful for skewed environmental data}.
\end{align*}

\subsection{Better Extreme Event Prediction}
\begin{itemize}
    \item Standard models underestimate tail risks \cite{Williams2020}.
    \item The $q$-Gamma distribution is taken to better capture both rare low- ($x\to 0$) and high-pollution ($x \to \infty)$ events:
    \[
    P(x) \propto x^{\alpha-1} \left[ 1 + (q-1)\lambda x \right]^{\frac{1}{1-q}}
    \]
\end{itemize}

\subsection{Moments of the Distribution}

\subsubsection{Mean Concentration}
\[
\langle x \rangle = \frac{1}{Z} \int_0^\infty x^\alpha [1 + (q-1)\lambda x]^{\frac{1}{1-q}} \, dx,
\]
\[
\langle x \rangle = \frac{[(q-1)\lambda]^{\alpha} \, \Gamma\left( \frac{1}{q-1} \right)}{\Gamma(\alpha) \, \Gamma\left( \frac{1}{q-1} - \alpha \right)}
\frac{\Gamma(\alpha+1) \, \Gamma\left( \frac{1}{q-1} - \alpha-1 \right)}{[(q-1)\lambda]^{\alpha+1} \, \Gamma\left( \frac{1}{q-1} \right)}
\]
\[
\langle x \rangle = \frac{\alpha}{\lambda \left[ 1 - (\alpha+1)(q-1) \right]}
\]

\subsubsection{Variance}
\[
\mathrm{Var}(x) = \frac{1}{Z} \int_0^\infty x^{\alpha + 2} [1 + (q-1)\lambda x]^{\frac{1}{1-q}} \, dx -\langle x \rangle^{2},
\]
\[
\mathrm{Var}(x) = \frac{[(q-1)\lambda]^{\alpha} \, \Gamma\left( \frac{1}{q-1} \right)}{\Gamma(\alpha) \, \Gamma\left( \frac{1}{q-1} - \alpha \right)}
\frac{\Gamma(\alpha+2) \, \Gamma\left( \frac{1}{q-1} - \alpha-2 \right)}{[(q-1)\lambda]^{\alpha+2} \, \Gamma\left( \frac{1}{q-1} \right)}-\frac{\alpha^2}{\lambda^2 \left[ 1 - (\alpha+1)(q-1) \right]^2}
\]
\[
\mathrm{Var}(x) = \frac{\alpha(2-q)}{\lambda^2 \left[ 1 - (\alpha+1)(q-1) \right]^2 \left[ 1 - (\alpha+2)(q-1) \right]}
\]

\subsection{Interpretation}
This distribution combines:
\begin{itemize}
    \item \textbf{Power-law term} ($x^\alpha$) for low concentrations $x$,
    \item \textbf{q-exponential term} for heavy-tailed extreme events, leading to asymptotic power law $x^{-\frac{1}{q-1}}$ for $x \to \infty$.
\end{itemize}

Previous work \cite{He2022} only fitted the tails of the data by $q$-exponentials. We now look at the entire distribution. 

\begin{table}[h]
\centering
\caption{Physical Interpretation of Parameters}
\label{tab:params}
\begin{tabular}{ll}
\toprule
Parameter & Physical Meaning \\
\midrule
$\alpha$ & Shape at low concentrations \\
$q$ & Non-extensiveness \\
$\lambda$ & Decay rate \\
\bottomrule
\end{tabular}
\end{table}

\subsection{Maximum Likelihood Estimation (MLE)}
\textbf{Maximum Likelihood Estimation (MLE)} is a fundamental method for estimating the parameters of a statistical model\cite{Casella2002}. The principle is to find the parameter values that maximize the \textbf{likelihood function} $\mathcal{L}(\boldsymbol{\theta} | \mathbf{x})$, which represents the probability of observing the given sample data $\mathbf{x} = (x_1, x_2, \dots, x_n)$.

\begin{enumerate}
    \item \textbf{Likelihood Function:} Assuming independent and identically distributed (i.i.d.) data, the joint likelihood for the $q$-Gamma distribution is the product of the individual probability densities:
    \[
    \mathcal{L}(\alpha, q, \lambda | \mathbf{x}) = \prod_{i=1}^{n} f(x_i | \alpha, q, \lambda) = \prod_{i=1}^{n} \frac{1}{Z} \frac{x_i^{\alpha-1}}{\left[1 + (q - 1)\lambda x_i\right]^{\frac{1}{q-1}}}
    \]
    where $Z$ is the normalization constant.

    \item \textbf{Log-Likelihood:} Maximizing the product in the above equation is numerically unstable. We instead maximize the \textbf{log-likelihood}, which converts the product into a sum:
    \[
    \ell(\alpha, q, \lambda) = \log \mathcal{L}(\alpha, q, \lambda | \mathbf{x}) = \sum_{i=1}^{n} \log f(x_i | \alpha, q, \lambda) \\
    \]

\end{enumerate}

\subsection{Sum of Squared Residuals (SSR)}

The \textbf{Sum of Squared Residuals (SSR)} quantifies the discrepancy between the observed data and the values expected under the fitted model. Minimizing the SSR achieves the best possible fit of a model to the data\cite{Seber2003}.

\noindent \textbf{Implementation:}

\begin{enumerate}
    \item \textbf{Binning:} The data is grouped into $K$ bins (histogram) with observed frequencies $O_k$ for $k = 1, \dots, K$.

    \item \textbf{Expected Frequencies:} Using the fitted PDF with MLE parameters $(\hat{\alpha}, \hat{q}, \hat{\lambda})$, the expected frequency $E_k$ for bin $k$ is:
    \[
    E_k = n \int_{\text{lower}_k}^{\text{upper}_k} f(x | \hat{\alpha}, \hat{q}, \hat{\lambda})  dx
    \]
    where $n$ is the total number of observations.

    \item \textbf{Calculation of SSR:} The SSR is computed as the sum of squared differences between observed and expected bin counts:
    \[
    \text{SSR} = \sum_{k=1}^{K} (O_k - E_k)^2
    \]
\end{enumerate}

\subsection{Log-Likelihood Goodness-of-Fit}

The \textbf{log-likelihood} provides a direct measure of model fit. A higher value indicates that the fitted distribution provides a better fit to the observed data\cite{Burnham2002}.

\noindent \textbf{Implementation:}

\begin{enumerate}
    \item \textbf{Model Fit:} The maximized log-likelihood is
    \begin{equation*}
    \ell(\hat{\alpha}, \hat{q}, \hat{\lambda}) = \sum_{i=1}^{n} \log f(x_i | \hat{\alpha}, \hat{q}, \hat{\lambda}).
    \end{equation*}

    \item \textbf{Model Comparison:} Between competing models, the one with the higher maximized log-likelihood is preferred.

    \item \textbf{Penalized Criteria:} To balance fit and complexity, criteria such as
    \begin{equation*}
    \text{AIC} = -2 \ell + 2k, \quad \text{BIC} = -2 \ell + k \log(n)
    \end{equation*}
    are commonly used, where $k$ is the number of parameters.
\end{enumerate}

\subsection{Anderson-Darling Goodness-of-Fit}

The \textbf{Anderson-Darling (AD) statistic} evaluates how well a model's cumulative distribution function (CDF) fits the observed data, placing greater emphasis on the tails\cite{Anderson1954}.

\noindent \textbf{Implementation:}

\begin{enumerate}
    \item \textbf{Empirical CDF:} Let $F_n(x)$ be the empirical CDF of the sample, and $F(x)$ the CDF of the fitted model.

    \item \textbf{AD Statistic:} Compute
    \begin{equation*}
    A^2 = -n - \frac{1}{n} \sum_{i=1}^{n} \Big[ (2i-1) \log F(x_{(i)}) + (2n+1-2i) \log (1-F(x_{(i)})) \Big],
    \end{equation*}
    where $x_{(i)}$ are the ordered sample values.

    \item \textbf{Interpretation:} Smaller $A^2$ indicates a better fit. It can be used to compare different models or check against critical values for hypothesis testing.
\end{enumerate}

\section*{Data Availability}
The data used in this study are publicly available at \url{https://uk-air.defra.gov.uk/data/data_selector_service?=&1=&s=1&o=#mid}. The code used for fitting and generating plots are available at \url{https://github.com/CptMcSalad/Superstatistical-Analysis-of-PDFs}. 

\section*{Acknowledgements}
C.B. acknowledges funding by a QMUL ISPF-ODA Research England grant on air pollution dynamics,  as well as funding by STFC grant UKRI467. 

\section*{Author contribution}
Conceptualization - All authors; Methodology - All authors; Data curation - All authors. Data visualization -
All authors; Writing original draft - All authors. All authors approved the final submitted draft.

\section*{Competing interest.}
The authors declare no competing interests.

\end{document}